\documentclass{aa}
\usepackage{graphicx,lscape,rotating,psfig}

\newcommand{\lapprox}{$_ <\atop{^\sim}$} 

\begin{document}

\title{Discovery of Radio-loud Quasars with Redshifts above 4 from the PMN sample}

   \subtitle{}

\titlerunning{New Radio-loud Quasars with Redshifts above 4}

\author{Isobel M. Hook\inst{1}\thanks{\emph{Present address:} Gemini Observatory, 670 N. A'ohoku Place, Hilo, HI 96720, USA.}
\and 
Richard G. McMahon\inst{2} 
\and 
Peter A. Shaver\inst{3} 
\and 
Ignas A. G. Snellen\inst{4}}

\offprints{I. M. Hook}

\institute{
 Department of Physics, University of Oxford,
 Nuclear \& Astrophysics Laboratory,
 Keble Road,
 Oxford,  OX1 3RH, UK\\
 \email{imh@astro.ox.ac.uk} 
\and 
Institute of Astronomy, Madingley Road, Cambridge CB3 0HA, UK \and
European Southern Observatory, Karl Schwarzschild Stra\ss{e} 2, 
    D-85748 Garching b. M\"{u}nchen, Germany 
\and
Institute for Astronomy, Royal Observatory, Edinburgh, EH9 3HJ, UK}

\date{version \today}

\abstract{\thanks{The work presented here is based in part on data obtained with the ESO facilities on La Silla (EFOSC/3.6-m)}\ \thanks{Figures 7 and 8 of this paper are available only in the electronic Journal.}
We present results of a new, large survey for high-redshift radio-loud
quasars, which targets quasars with $z>4$.  The survey is based on the
PMN and NVSS radio surveys, optically identified using digitised UKST
B, R and I plates.  Six new $z>4$ flat-spectrum QSOs have been
discovered, and one previously known $z>4$ QSO rediscovered, based on
their red optical colours.  The QSOs discovered in this survey are
bright in both radio and optical bands; in particular PMN J1451-1512
($z=4.763$, $\rm I=17.3$, $R=19.1$) and PMN J0324-2918 ($z=4.630$, $\rm
R=18.7$) are very luminous.  PMN J1451-1512 at $z=4.763$ is also now
the most distant radio-selected quasar.  In addition, 9 new quasars
with $3.5<z<4.0$ were discovered during the survey. We present spectra
and finding charts for the new quasars. We also derive a surface
density of $\rm 1.0\pm 0.4\times 10^{-3} sq\ deg^{-1}$ for
flat-spectrum $z>4$ QSOs with $\rm S\ge 72$mJy and $\rm R< 21$mag.

}

\maketitle

\section{Introduction}

Radio selection remains one of the most efficient ways of finding
high-redshift AGN. This approach has the further advantage of being
less prone to selection effects than optical selection, since radio
emission is unaffected by either intrinsic or extrinsic absorption due
to dust. The specific aim of this work was to find optically bright,
radio-selected high-redshift quasars. These can be used for unbiased
studies of damped Lyman alpha systems at high redshift and other
follow-up studies such as searches for associated high-redshift galaxy
clusters.

We therefore began to carry out a large, systematic survey aimed
specifically at $z>4$ QSOs. Our method involves the optical
identification of flat-spectrum radio sources and the spectroscopic
follow-up of the red stellar identifications. This approach exploits
the fact that quasars at high redshift have redder optical colours
than their low-redshift counterparts due to absorption by intervening
HI (see figure 1 in Hook {\it et al.}  1995), and has proved
successful at finding high-redshift quasars in the past (Hook et al
1995, 1996, 1998).

Previous work using well-defined quasar samples has shown that $z>4$
radio-loud quasars are likely to be rare objects, both because the
quasar population as a whole appears to decline at redshifts above
$\sim 2-3$ (Kennefick et al. 1996; Hawkins \& Veron 1996, Schmidt,
Schneider \& Gunn 1995; Warren, Hewett \& Osmer 1995) and because
radio-loud quasars represent only about 10\% of the full quasar
population. 

Specific studies of the radio-loud quasar population have shown that
these objects are indeed rare at $z>4$.  Dunlop \& Peacock (1990)
presented strong evidence for a drop in the space density of
radio-loud quasars between $z=2$ and $z\approx 3$ based on
radio-selected samples reaching $\rm S_{2.7GHz} = 100mJy$.  More
recently, significant progress has been made towards understanding the
evolution of the radio-loud quasar population out to $z\sim 4$ and the
potential effects of absorption by dust, by the study of a completely
identified, large area, flat-spectrum radio sample with $\rm
S_{2.7GHz}\ge250mJy$ (Shaver et al. 1996; Wall et al., in
preparation).  The low numbers of high-reshift quasars found in these
studies demonstrates that there is a distinct drop-off in the space
density of quasars at $z>3$. Thus for our new survey to be successful
it must reach fainter radio flux density limits than the above surveys
(to sample further down the luminosity function), and cover a
significant fraction of the sky.
 
Here we present the first results of this new survey for high-redshift
radio-loud quasars. As will be seen in section 2, the survey uses
deeper radio and optical data than previous radio-loud quasar surveys,
and covers a very large area in the Southern sky ($\sim 7,500$ sq deg,
comparable to that of the planned 10,000 sq deg of the Sloan survey).
Our survey has produced seven $z>4$ flat-spectrum quasars, one of
which was previously known. The survey complements the survey of
Snellen et al (2001) which contains four flat-spectrum $z>4$ in the
Northern sky, selected using a a similar method.

\section{The Radio Sample and the Optical Identification procedure}

The parent radio sample used in this study is based on the
Parkes-MIT-NRAO radio survey (PMN, Griffith et al 1995 and references
therin), selected at 5GHz. The data cover the southern sky with
$\delta < 10^\circ$ to a flux density limit of 20-72mJy depending on
declination. No additional flux density limit was applied when
carrying out the survey, but note that when considering the statistics
of the final quasar sample, we consider sources with $\rm S\ge 72mJy$,
since the PMN completeness at lower flux density levels is patchy (see
Figure~\ref{aitrad}).

To provide accurate positions and spectral index information, the PMN
catalogue was matched to the 1.4GHz NRAO-VLA Sky Survey (NVSS, Condon
et al. 1998), which covers the declination range $\delta \ge
-40^\circ$.  All NVSS sources matching within a 2 arcminute radius of a PMN
position were kept.

Since the beam sizes of the PMN and NVSS surveys are different ($4.2'$
for PMN and $45''$ for NVSS), the 1.4 flux density used to determine
spectral indices were calculated by summing the total flux density
from all NVSS sources whose positions were within a 2 arcminute radius
of the PMN position.  Flat spectrum objects with $\rm S_{PMN} \ge
50mJy$ were then selected using the criterion $\alpha^{5}_{1.4} \ge
-0.5$ where $S\propto\nu^{\alpha}$. This gave 5976 PMN radio sources.

The NVSS positions were matched to optical catalogs which were derived
from scans of UK Schmidt Telescope (UKST) plates. The scans were
produced using the Automatic Plate Measuring (APM) facility (McMahon
\& Irwin 1992).  The optical identification procedure was similar to that
used in the past to make POSS-based identifications of radio samples,
described in Hook et al (1995,1996,1998). However the current survey
also makes use of I-band plates for the first time.

The plate data used is from the UKST survey of the southern sky in the
B, R and I bands, reaching limiting magnitudes of approximately 22.6,
21.0 and 19.5 mag respectively. The blue plates are in the $\rm B_J$
passband, $\rm 3950-5400\AA$, and have 606 centres spaced at
intervals of 5 degrees. In most cases the matching R plate used was
from the OR survey with a passband of $\rm 5950-6900\AA$, although in
some fields not yet covered by the OR survey, the the old R plate was
used ($\rm 6300-6900\AA$). The I plates used IV-N emulsion + RG715
filter giving a passband of $\rm 7150-8900\AA$.

The maximum redshift of quasars that our survey can detect in
principle is defined by the redshift at which the $\rm Ly-\alpha$ line is
redshifted out of the redder passbands.
For the R-plates this occurs at redshifts above $z=4.7$ and for the I
plates at redshifts above $z=6.3$.

\subsection{Area covered by the survey}

\begin{figure*} 
\centerline{\psfig{figure=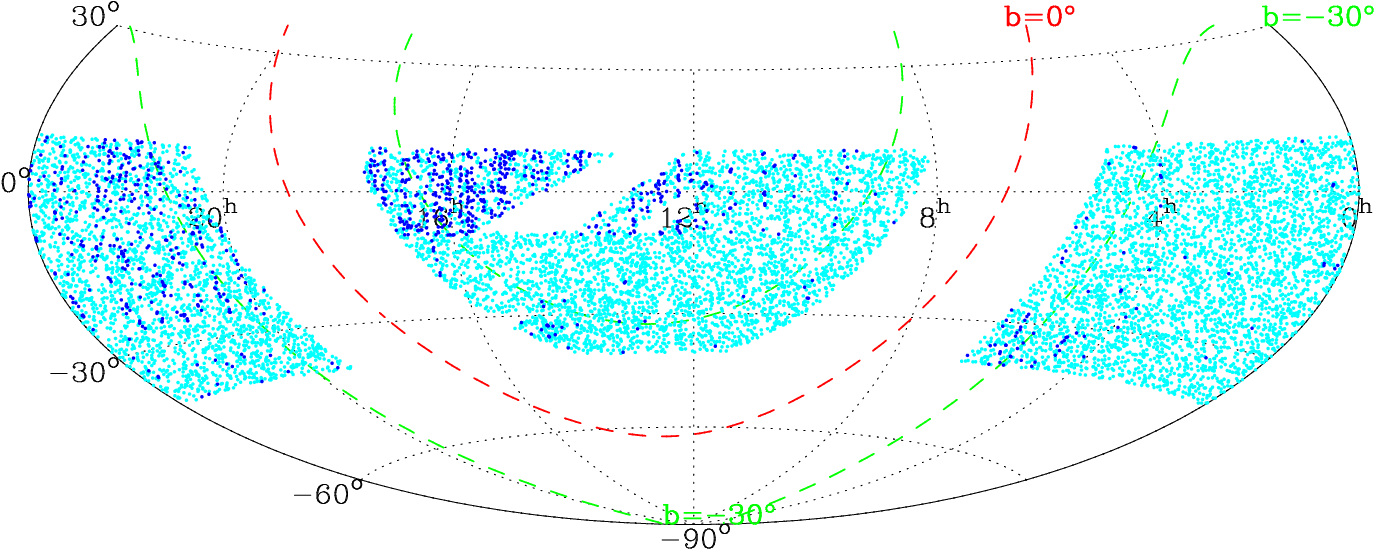,height=2.5in,angle=270.0}}
\caption{Area covered by the radio data at the time of the survey. 
light dots show PMN sources with $\rm S\ge 72mJy$ which had a
counterpart in the NVSS catalogue, and dark dots show those PMN
sources which did not have an NVSS counterpart. Usually this was
because the NVSS survey had not been completed in those areas at the
time (March 1998).}
\label{aitrad}

\centerline{\psfig{figure=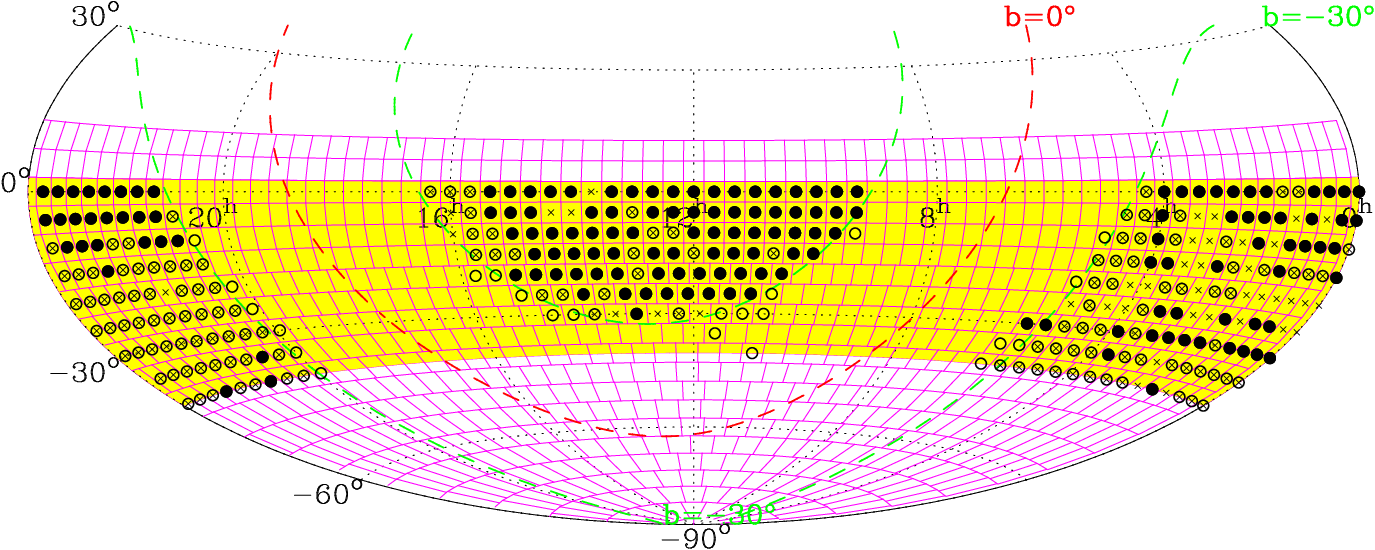,height=2.5in,angle=270.0}}
\caption{Area covered by the optical data. Crosses show UKST fields whose
centres satisfy $|b|>30\deg$, $-40\deg<\delta<0\deg$, open circles
show fields that have B and R plates, and filled circles show fields
that have B, R and I plates. A total of 313 fields have at least B and
R plates available, corresponding to an area of 7525sq deg. The shaded
area represents the declination range of the survey, defined by the
overlap of the NVSS, PMN and UKST data.}
\label{aitsky}
\end{figure*}

Figure~\ref{aitrad} shows the spatial distribution of the radio
sources which form the basis of the survey. These lie in the region of
overlap of the PMN and NVSS surveys. The effective area of our quasar
survey is defined by the subset of this region that is covered by APM
scans of UKST plates.

All the available B, R and I plates have now been scanned in the
region of overlap with our radio sample ($-40^{\circ} < \delta <
+2.5^{\circ}$, avoiding the galactic plane, $|b| > 30^{\circ}$).  At
the time of the spectroscopic observations reported here, an area of
7525 sq degrees was covered by plates in the B and R bands (4637.5sq
deg in the South Galactic Cap region and 2887.5sq deg in the North
Galactic Cap region). A total of 3887.5 sq degrees also had I-band
plates.  The spatial distribution of the plates is shown in
Figure~\ref{aitsky}. Of this, two regions around RA=13h,
DEC=0$^{\circ}$ have no radio data, as shown in Figure~\ref{aitrad}.
The total area of overlap with the B,R data is 7265.5 sq deg of which
4637.5sq deg is in the South Galactic Cap (SGC) region and 2628.0 sq
deg is in the North Galactic Cap (NGC). Of this a total of 3637.5 sq
deg also had I data.

In addition there are small areas within the region of radio/optical
overlap that were not yet covered by the NVSS survey at the time our
QSO survey was carried out. This resulted in some PMN sources not
being matched with an NVSS counterpart in certain regions, as can be
seen in figure~\ref{aitsky}. When considering the statistics of our
final quasar sample, we take this incompleteness into account in a
statistical way by calculating the fraction of PMN sources with $\rm
S_{PMN}>72mJy$ that do not have NVSS counterparts. In the SGC the
matched fraction is 0.955 and in the NGC it is 0.871.

\subsection{Selection of the spectroscopic sample}

\begin{figure} 
\centerline{\psfig{figure=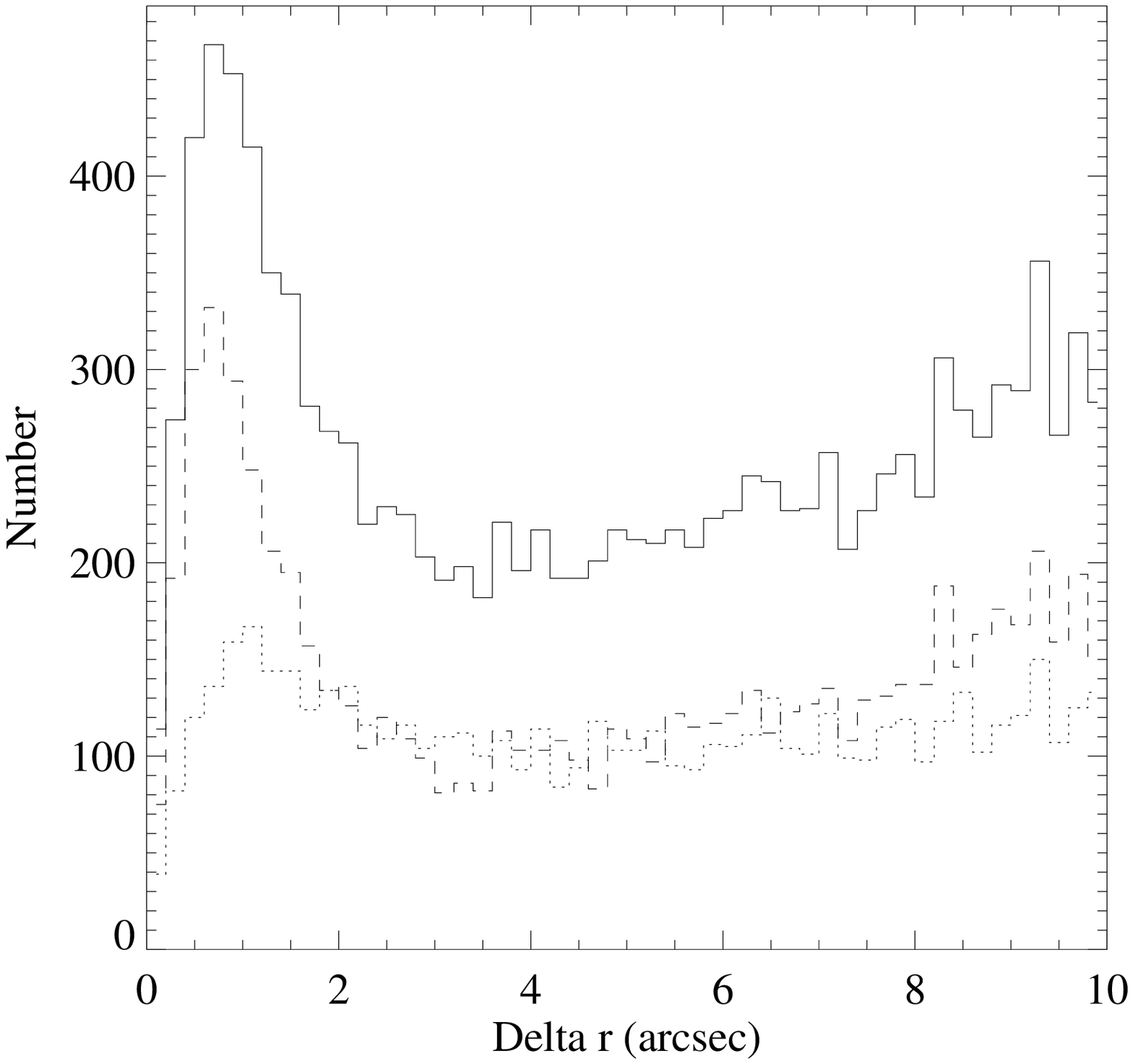,height=2.8in,bbllx=97pt,bblly=18pt,bburx=574pt,bbury=476pt}}

\caption{Distribution of differences between the optical (APM) and radio (NVSS) positions. The solid line is the histogram for all sources, the dashed line is for objects classified as stars and the dotted line is for objects classified as galaxies. A matching radius of 3 arcsec was used in this survey.}
\label{radplot}
\end{figure}

Optical identifications were made based on positional coincidence of
the NVSS position with an optical counterpart on the UKST plates. A
matching radius of $3.0''$ between the NVSS and optical positions was
used (see Figure~\ref{radplot} for the distribution of positional
differences).

\begin{figure*} 
\centerline{\psfig{figure=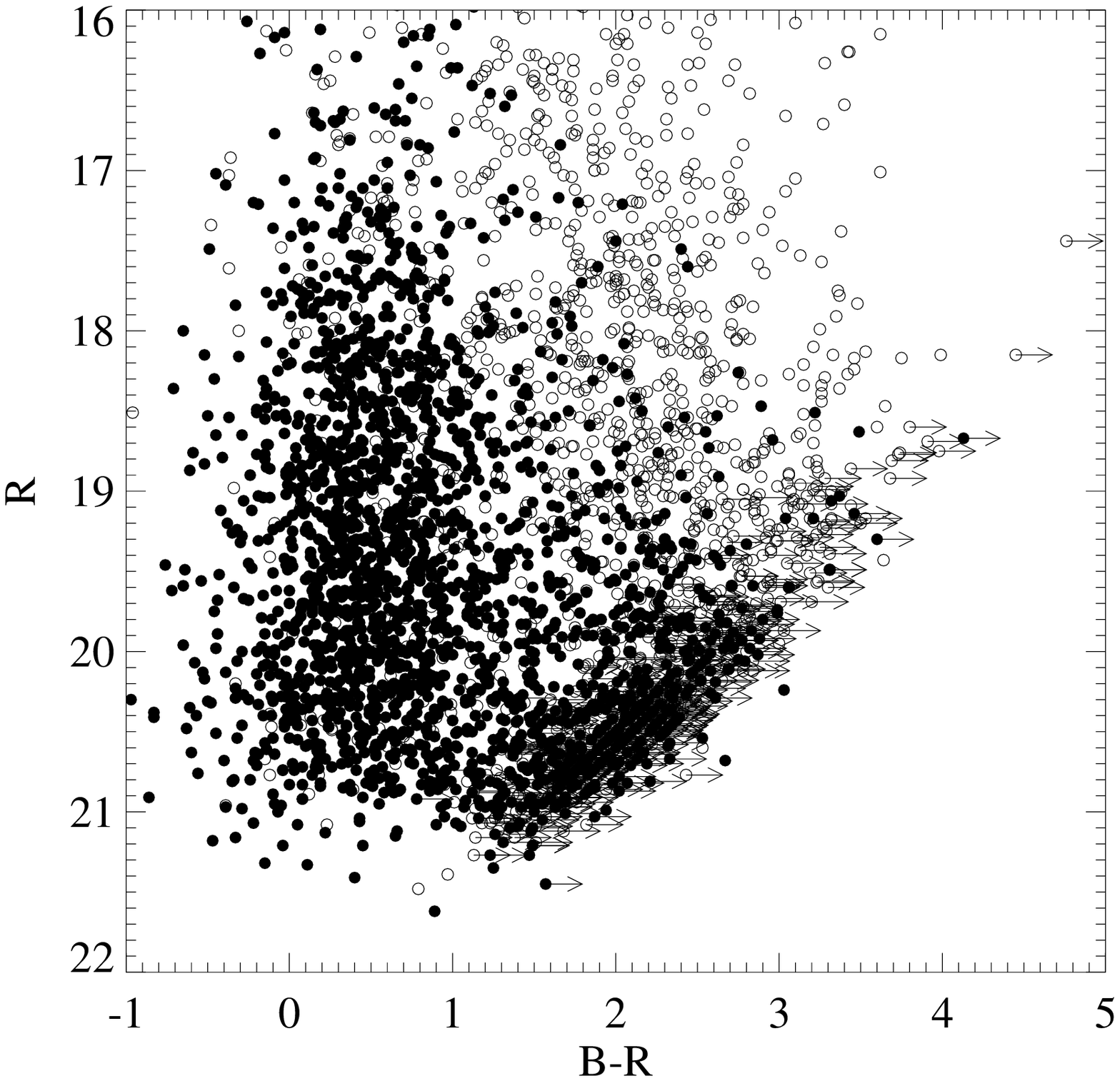,height=2.6in,bbllx=97pt,bblly=18pt,bburx=574pt,bbury=476pt}\hspace{0.25cm}\psfig{figure=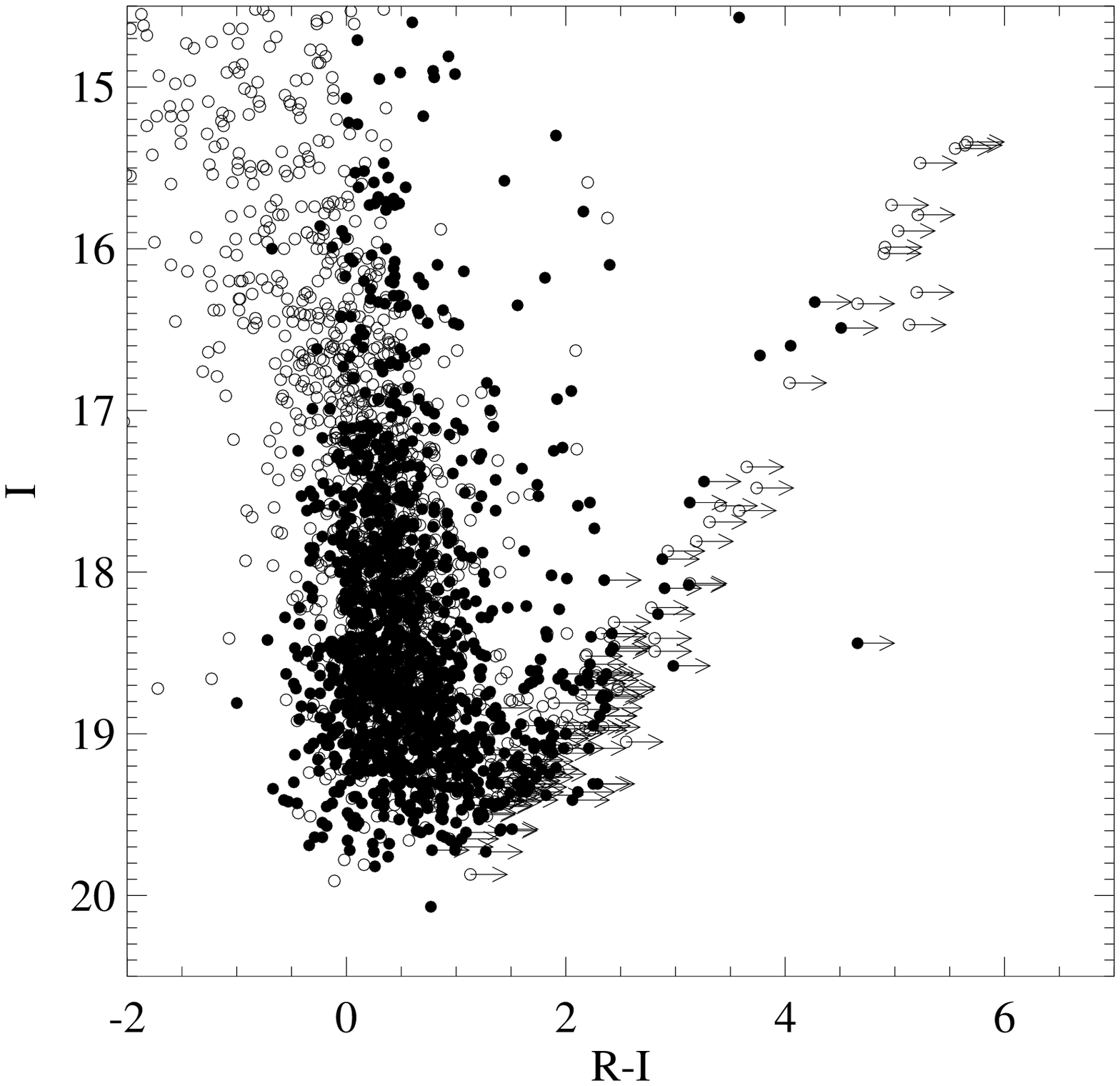,height=2.6in,bbllx=97pt,bblly=18pt,bburx=574pt,bbury=476pt}}
\vspace{0.25cm}
\centerline{\psfig{figure=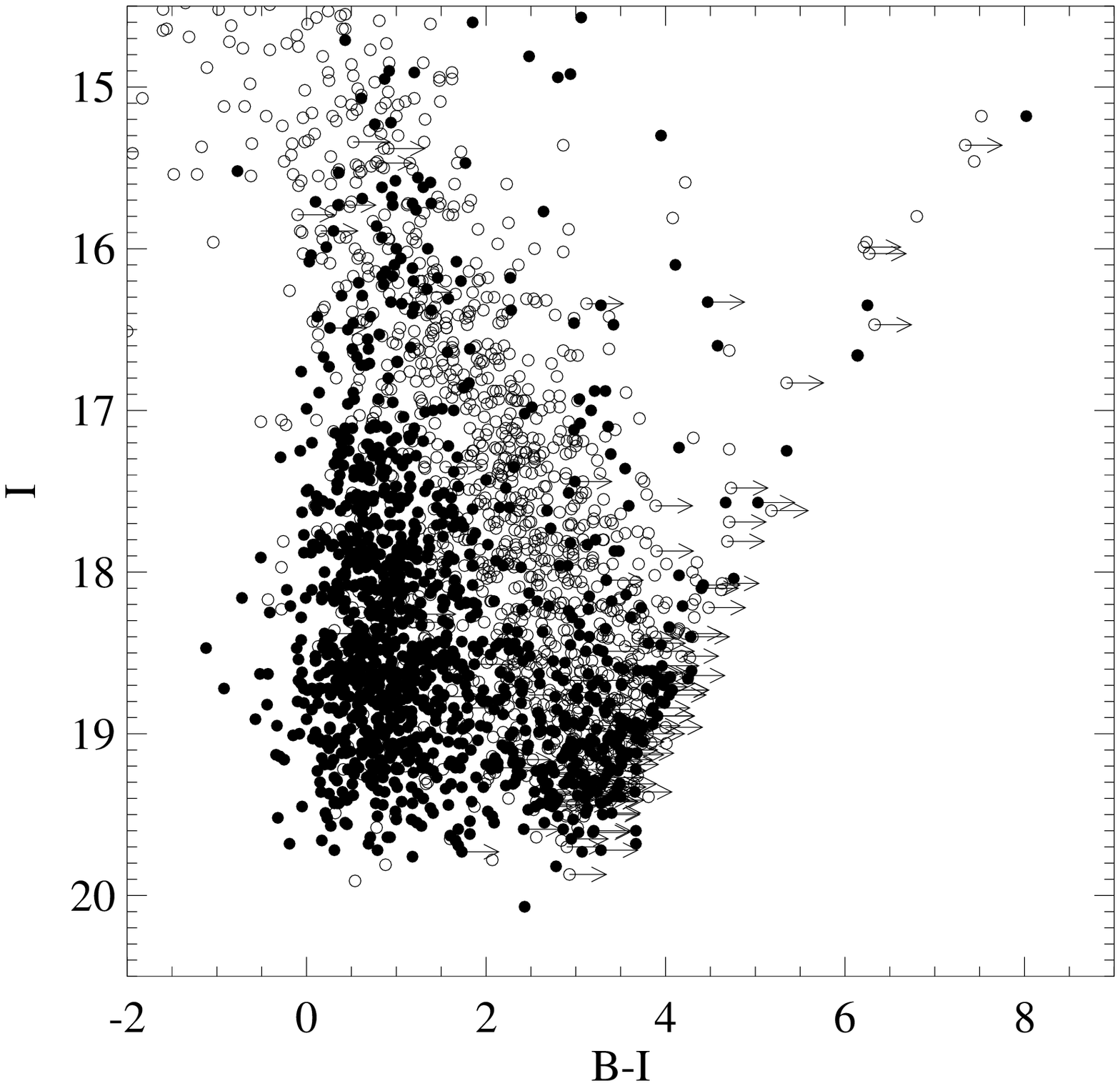,height=2.6in,bbllx=97pt,bblly=18pt,bburx=574pt,bbury=476pt}\hspace{0.25cm}\psfig{figure=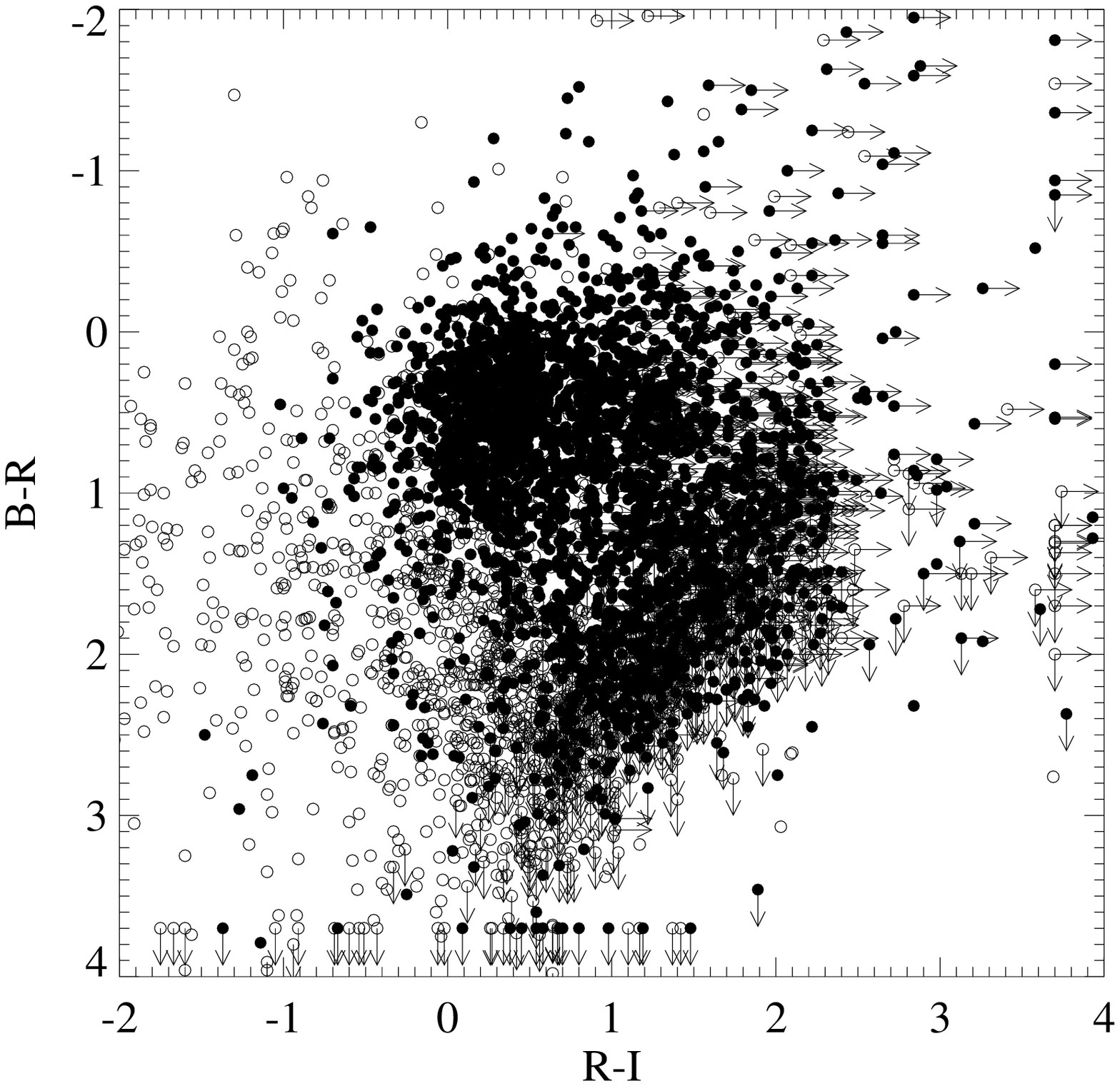,height=2.6in,bbllx=97pt,bblly=18pt,bburx=574pt,bbury=476pt}}
\caption{Colour-magnitude and Colour-colour plots for the APM identifications within 3$''$ of the NVSS radio position. Solid symbols represent objects classified as stellar, open symbols represent objects classified as galaxies by the APM. Arrows are used to represent limits on the colours when objects are not detected in one or more of the three passbands.}
\label{colmag}
\end{figure*}

From these identifications, red, stellar objects were selected for
spectroscopic follow up. Figure~\ref{colmag} shows colour-magnitude
and colour-colour diagrams for optical identifications within $3''$ of
the NVSS position. The first spectroscopic sample contained any
stellar object with $\rm B-R\ge 1.5$. A second sample, the `I-band
sample', was then defined, which contained stellar objects with $\rm
R-I \ge 1.0$, or $\rm B-I \ge 2.0$ that were not already included in
the first sample.  The criterion for being considered a stellar object
was $\sigma_{class} \le 3.0$ where $\sigma_{class}$ is the APM
classification parameter measured from the R plate. If the object was
not detected in R then the I plate classification was used. These
selection criteria resulted in a sample of 228 sources. Of these, 33
had known redshifts from the literature prior to the start of this
project.

\section{Spectroscopic observations and results}

\vspace{0.4cm} In total, 169 candidates were observed
spectroscopically.  During two runs in October 1998 and May 1999, 157
objects were observed at the ESO 3.6m, La Silla, Chile using the EFOSC
spectrograph. For both runs the detector used was 'ccd40', a Loral CCD
with $2048\times2048$ pixels, binned $2\times2$. For the 1999 May
observations grism \#12 was used, covering the spectral range
5800\AA--9500\AA with a dispersion of 4.23\AA\ per binned pixel. For
the 1998 October observations grisms \#4 and \#5 were used covering
the ranges 3420--7460\AA\ and 6000--10000\AA\ respectively with dispersions of
4.08 and 4.27\AA\ per binned pixel. The slit width was chosen to match
the seeing, typically 1.2--1.5 arcsec.  For the 1998 October
observations a sky PA of $270^\circ$ was used, and for the 1999 May
observations the slit was at the paralactic angle. Approximate
relative flux calibration was obtained using observations of
spectrophotometric standards from Hamuy et al (1994).  Twelve sources
were observed at AAT using the RGO spectrograph in October 1996.

26 sources remain to be observed, of which 24 are in the North
Galactic Cap (NGC) and 2 are in the South Galactic Cap (SGC) region.

\begin{table*}
\caption{Summary of current survey status for sources with $\rm S\ge 50mJy$.}
\begin{tabular}{|l r| c c c|}\hline
\multicolumn{2}{|c|}{Number of candidates} & \multicolumn{3}{c|}{Number confirmed}\cr 
\multicolumn{2}{|c|}{with $z$ data} & $3.5<z<4.0$ &  $4.0<z<4.5$ & $z>4.5$ \cr\hline
Present observations  & 169     &  7      &  4     & 2 \cr
Literature data       &  33     &  2      &  1     & 0 \cr\hline
Total                 & 202     &  9      &  5     & 2 \cr\hline
\end{tabular}
\label{status}
\end{table*}

The data were reduced using standard procedures with the
IRAF\footnote{IRAF is distributed by the National Optical Astronomy
Observatories, which is operated by the Association of Universities
for Research in Astronomy, Inc. (AURA) under cooperative agreement
with the National Science Foundation.} software environment.

\begin{figure*}
\centerline{\psfig{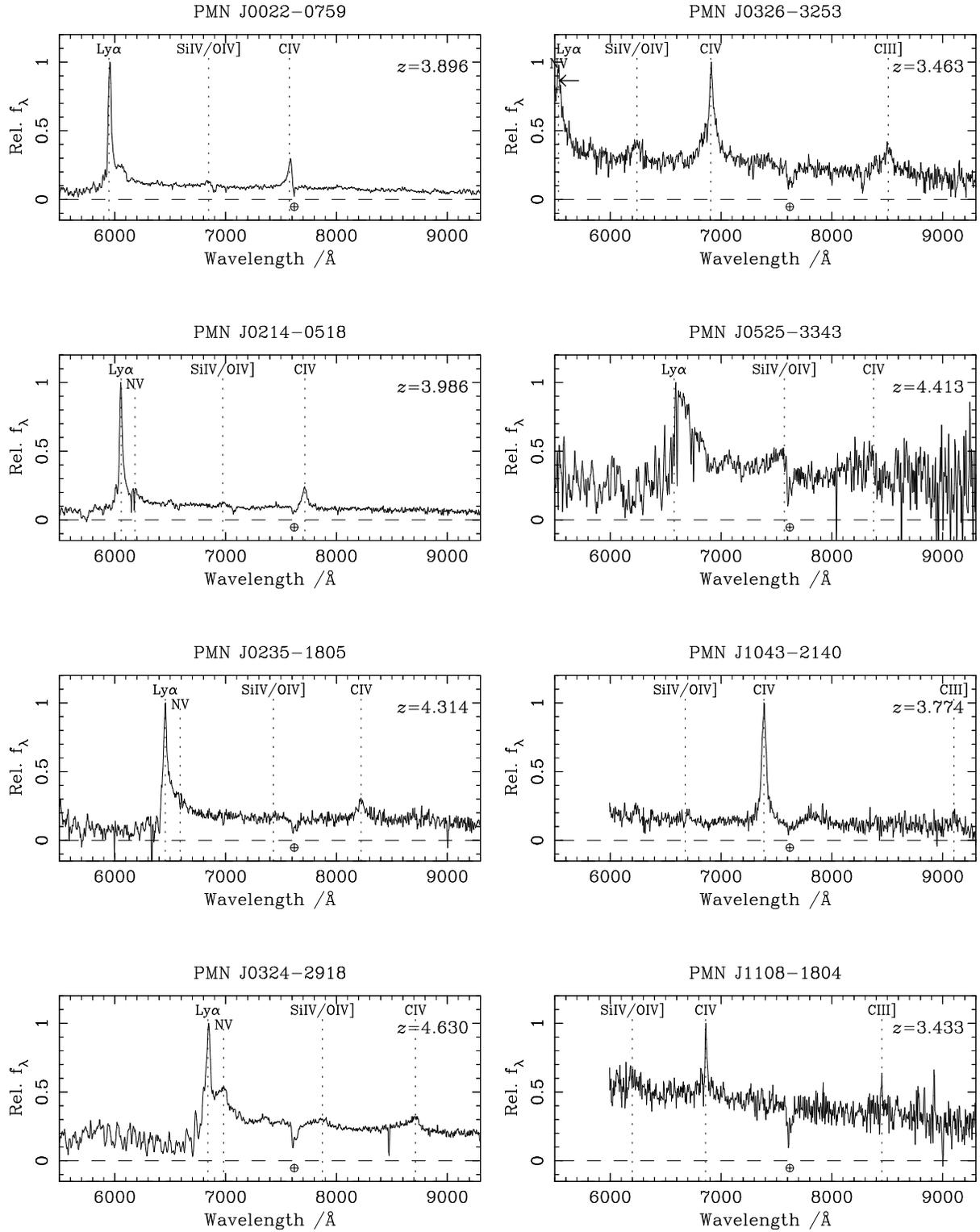}}
\caption{Optical spectra of the $z>3$ quasars.}
\label{spectra}
\end{figure*}

Table~\ref{status} summarises the current survey status and in
Table~\ref{ztab} we give redshifts, optical magnitudes and radio flux
densities for the objects which were identified as new high-redshift
quasars.  The spectra are shown in Fig.~\ref{spectra}, and finding
charts, with J2000 coordinates, are given in Figure~\ref{fct}.  Two of
the $z>4$ QSOs are discussed in more detail below.

\paragraph*{\bf PMNJ1451-1512 }

This is now the most distant radio-selected quasar. The object is
blank on the UKST B-band plate, detected in R, and very bright in I
(17.3). Near-IR magnitudes from the 2MASS survey are J =
$16.39\pm0.10$, $H = 15.28\pm 0.09$, $K = 14.67\pm 0.09$.

The optical spectrum (Fig.~\ref{spectra}) shows strong lines of $\rm
Ly\alpha$ SiIV/OIV and CIV from which we estimate a redshift of
4.763. However the absorption of the CIV and Ly-$\alpha$ line make
redshift determination uncertain.  A 4800s IR spectrum was obtained
using SOFI at the 3.5m NTT, La Silla on 31 May 1999. The spectrum
shows strong, broad emission lines of CIII and MgII at a redshift of
4.764 and approximately 4.78 respectively (the MgII line has a complex
shape, see Figure~\ref{irspec}). We adopt a redshift of 4.763 for this
QSO.

The NVSS map (see Fig.~\ref{nvss}) shows a source at the optical
position of the QSO, and a second NVSS source about 4 arcmin away. The
PMN position appears to lie between the two, although much closer to the
first position (that of the QSO).  It is possible that the PMN flux
density is overestimated because of a contribution from the second
source.

\paragraph*{\bf PMN J0525-3343 }
This QSO was discovered in the AAT run of October 1996. It was
reobserved at the Las Campanas 100'' du Pont telescope using the
ModSpec spectrograph and this spectrum is shown Figure~\ref{spectra}.
The somewhat lower signal-to-noise combined with the lack of clean
strong lines make the redshift determination difficult. However
P\'eroux et al (2001) have since obtained a higher quality spectrum
from which a more accurate redshift measurement can be made. They
obtain $z=4.383\pm 0.034$ assuming rest wavelengths for the emission
lines that have not been corrected for systematic shifts, or $z=4.338$
when assuming the rest wavelengths used in Table~\ref{wave}, which
have been corrected for systematic shifts as described by Tytler \&
Fan (1992).

This source has been confirmed as an X-ray source with ROSAT HRI and
its X-ray properties are the subject of another paper (Fabian et al
2001).

\begin{sidewaystable*}
\caption{Summary of optical and radio properties for quasars with $z>3$ in the sample. A $`-'$ indicates that no data were available. The positions are the APM optical positions determined from the R plates.}
\begin{tabular}{lllccrrrrll}\hline
 Name             &\multicolumn{1}{c}{RA} &\multicolumn{1}{c}{DEC} & $z$ &\multicolumn{1}{c}{R} &\multicolumn{1}{c}{B}&\multicolumn{1}{c}{I}& $\rm S_{5GHz}$&\multicolumn{1}{c}{$\rm \alpha_{5GHz}^{1.4GHz}$} & Place \& Date & comments \cr 
                   & \multicolumn{2}{c}{J2000}      &        &        &      &   & &        & discovered& \cr\hline
 PMN J0022$-$0759 & 00 22 00.246 & $-07$ 59 16.03 & 3.896 & 19.47 & 21.26  &$>19.25$&   54 & $-0.18$  & ESO 10/98 &\cr
 PMN J0214$-$0518 & 02 14 29.295 & $-05$ 17 44.55 & 3.986 & 18.42 & 20.54  &  18.75 &   93 & $ 0.61$  & ESO 11/00 &\cr
 PMN J0235$-$1805 & 02 34 55.143 & $-18$ 06 08.49 & 4.314 & 18.79 & 22.00  &  $-$   &   49 & $-0.16$  & ESO 11/00 & $\rm S_{5GHz} < 50mJy$ (confused; see NVSS) \cr
 PMN J0324$-$2918 & 03 24 44.280 & $-29$ 18 21.10 & 4.630 & 18.66 & 22.18  &  $-$   &  354 & $ 0.30$  & ESO 11/00 &\cr
 PMN J0326$-$3253 & 03 27 00.407 & $-32$ 54 18.65 & 3.463 & 19.40 & 21.15  &  $-$   &   78 & $-0.01$  & ESO 11/00&\cr
 PMN J0525$-$3343 & 05 25 06.166 & $-33$ 43 05.34 & 4.413 & 18.50 & 21.47  &  $-$   &  210 & $ 0.06$  & AAT 10/96 &X-ray source, Fabian et al (2001)\cr 
 PMN J1043$-$2140 & 10 43 11.963 & $-21$ 40 47.97 & 3.774 & 20.66 &$>22.39$&$>18.99$&   59 & $-0.07$  & ESO 05/99 &\cr 
 PMN J1108$-$1804 & 11 08 48.035 & $-18$ 04 50.85 & 3.433 & 20.00 & 21.48  & 19.18  &   48 & $-0.31$  & ESO 05/99&  $\rm S_{5GHz} < 50mJy$\cr
 PMN J1429$-$1616 & 14 29 31.410 & $-16$ 15 40.44 & 3.842 & 19.76 & 21.26  & 19.42  &   50 & $ 0.34$  & ESO 05/99 &\cr
 PMN J1451$-$1512 & 14 51 47.052 & $-15$ 12 19.99 & 4.763 & 19.14 &$>22.60$& 17.25  &   90 & $ 0.89$  & ESO 05/99 & $\rm S_{5GHz}$ confused; see NVSS \cr
 PMN J2042$-$2223 & 20 42 57.278 & $-22$ 23 26.69 & 3.630 & 19.70 & 21.52  &  $-$   &  184 & $ 0.23$  & ESO 10/98&\cr
 PMN J2134$-$0419 & 21 34 12.006 & $-04$ 19 09.87 & 4.346 & 19.98 &$>22.80$& 19.73  &  221 & $-0.23$  & ESO 10/98&\cr
 PMN J2219$-$2719 & 22 19 35.304 & $-27$ 19 02.76 & 3.634 & 19.72 & 21.61  &  $-$   &  221 & $-0.27$  & ESO 11/00&\cr
 PMN J2220$-$3336 & 22 20 26.957 & $-33$ 36 59.44 & 3.691 & 21.08 & 22.73  &  $-$   &  123 & $ 0.67$  & ESO 11/00& $\rm S_{5GHz}$ confused; see NVSS \cr
 PMN J2314$+$0201 & 23 14 48.722 & $+02$ 01 50.86 & 4.110 & 19.87 &$>22.18$&$>19.28$&   97 & $-0.22$  & ESO 10/98& X-ray source, Boller et al (1997)\cr\hline
\multicolumn{8}{l}{Previously known QSOs contained in the sample}                 & & &Reference \cr\hline
 PMN J1028$-$0844 &              &                & 4.276 & 18.84 & 20.72  & 18.37  &  159 & $-0.43$  &           &Zickgraf et al (1997), X-ray source 0.3' away\cr
 PMN J1230$-$1139 &              &                & 3.528 & 19.45 & 21.95  &  $-$   &  374 & $-0.19$  &           &Drinkwater et al (1997)\cr
 PMN J2003$-$3251 &              &                & 3.78  & 17.60 & 19.31  &  $-$   & 1248 & $ 0.77$  &           &Peterson et al (1982) \cr\hline
\end{tabular}
\label{ztab}
\end{sidewaystable*}

\begin{table*}
\caption{Wavelengths of emission lines and redshifts for new $z>3$
objects, measured from the discovery spectra shown in figure
~\ref{spectra}. A $*$ indicates a line affected by absorption.
Redshifts are calculated assuming rest wavelengths corrected for
systematic shifts (Tytler \& Fan 1992) as follows: Ly-$\alpha$
1214.97\AA; CIV 1547.46\AA; CIII 1906.53\AA; SiIV/OIV
1398.62\AA. Uncertainties in the redshifts are approximately
$\pm$0.005, based on the range of redshift determined from different
emission lines in the same QSO.}

\begin{tabular}{llllll}\hline
 Name              & Ly$\alpha$ peak &  SiIV/OIV &  CIV  &  CIII      & mean $z$ \cr\hline
 PMN J0022$-$0759  & 5956.46       & 6831.46*&  7583.92*  & $-$       & 3.896 \cr 
 PMN J0214$-$0518  & 6053.03       & 6978.49 &  7714.46   & $-$       & 3.986 \cr 
 PMN J0235$-$1805  & 6455.15       & $-$     &  8225.35   & $-$       & 4.314 \cr 
 PMN J0324$-$2918  & 6846.70       & $-$     &  8703.26   & $-$       & 4.630 \cr 
 PMN J0326$-$3253  & 5422.49       & 6234.08 &  6913.65   & 8506.93   & 3.463 \cr 
 PMN J0525$-$3343  & 6627.50       & 7536.03 &  8347.96   & $-$       & 4.413$\dagger$ \cr 
 PMN J1043$-$2140  &  $-$          & $-$     &  7388.09   & 9101.68   & 3.774 \cr 
 PMN J1108$-$1804  & $-$           &  $-$    & 6863.86    & 8446.82   & 3.433 \cr 
 PMN J1429$-$1616  &  $-$          & 6770.20 &  7498.27   & 9229.22   & 3.842 \cr 
 PMN J1451$-$1512  & 7004.19       & 8057.75 &  8889.01*  &10989.2$\dagger\dagger$    & 4.763 \cr 
 PMN J2042$-$2223  & 5625.07       & $-$     &  7166.39   & $-$       & 3.630 \cr 
 PMN J2134$-$0419  & 6488.28       & 7488.82 &  8269.41   & $-$       & 4.346 \cr 
 PMN J2219$-$2719  & 5628.22       & 6482.09 &  7138.42*  & $-$       & 3.634 \cr 
 PMN J2220$-$3336  & 5684.32       & 6570.76 &  7265.08   & $-$       & 3.691 \cr 
 PMN J2314$+$0201  & 6195.83       & 7153.88 &  7914.37   & $-$       & 4.110 \cr 
\hline
\end{tabular}
\label{wave}
\newline
\noindent
$\dagger$ P\'eroux et al measure $z=4.388$ for this QSO (when using the above rest wavelengths) based on a higher-quality spectrum.\newline
$\dagger\dagger$ Line measured from IR spectrum
\end{table*}

\begin{figure*}
\centerline{\psfig{figure=MS2347f5b.ps,height=8.0in,bbllx=22pt,bblly=63pt,bburx=576pt,bbury=756pt}}
\end{figure*}

\begin{figure}
\centerline{\psfig{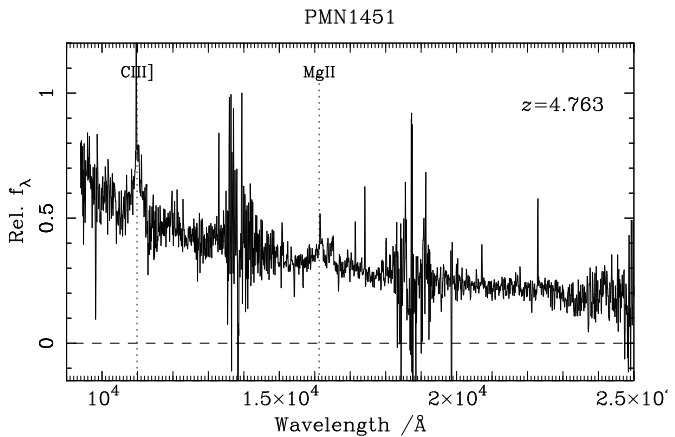}}
\caption{IR spectrum of PMN J1451-1512, obtained at NTT.}
\label{irspec}
\end{figure}

\begin{figure*}
\centerline{\psfig{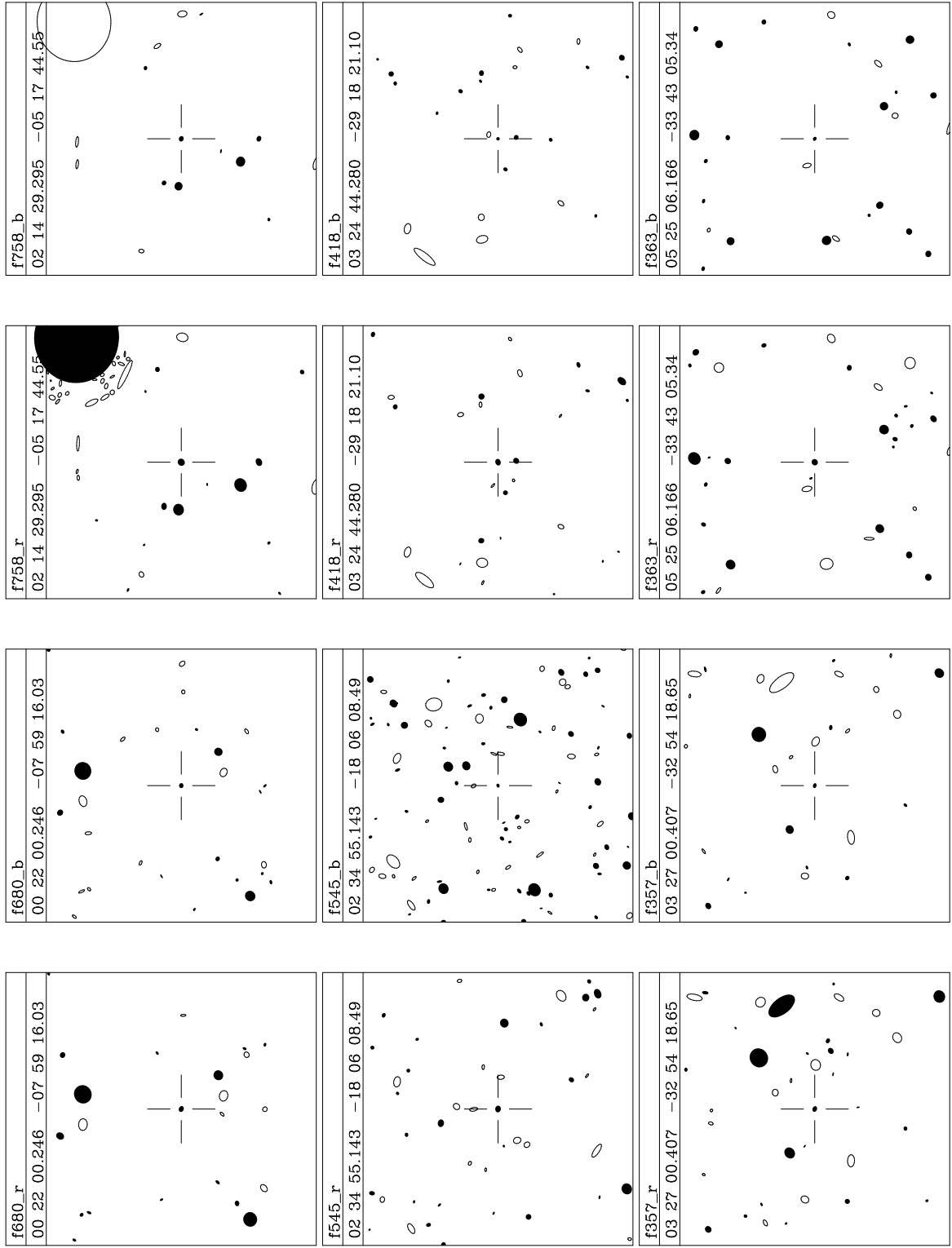}}
\caption{Optical Finding charts in B and R bands. J2000 positions are shown at the top of each chart. North
is to the top, East left.  These charts are from APM scans of UKST
plates. Filled symbols represent objects classified as stellar
by the APM, open symbols represent galaxies. The charts are 4 arcmin
across (central cross is 1 arcmin across), centered on the optical
position of the identification. }
\label{fct}
\end{figure*}

\begin{figure*}
\centerline{\psfig{figure=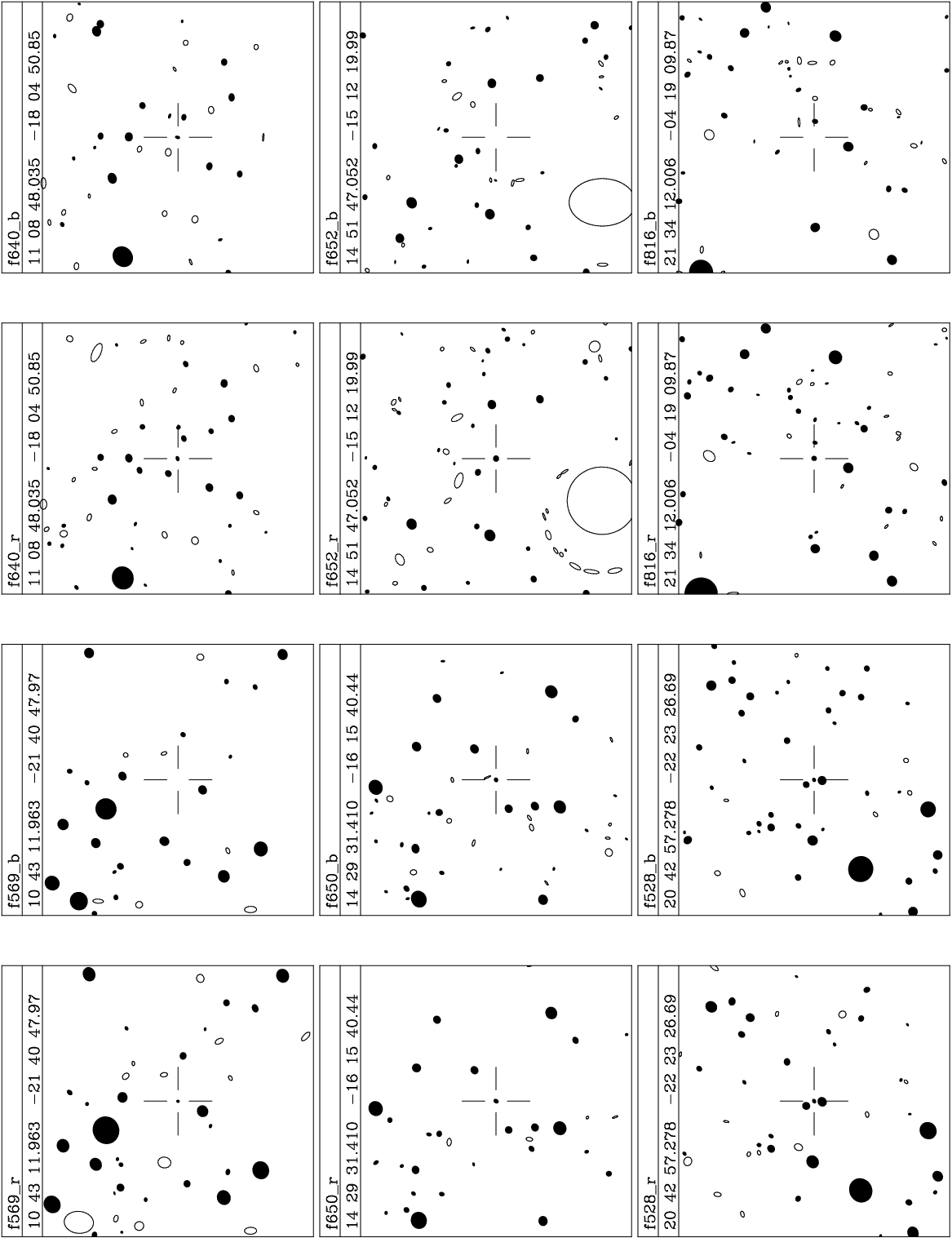,height=8.0in}}
\end{figure*}

\begin{figure*}
\centerline{\psfig{figure=MS2347f7c.ps,height=8.5in}}
\end{figure*}

\begin{figure*} 
\centerline{\psfig{figure=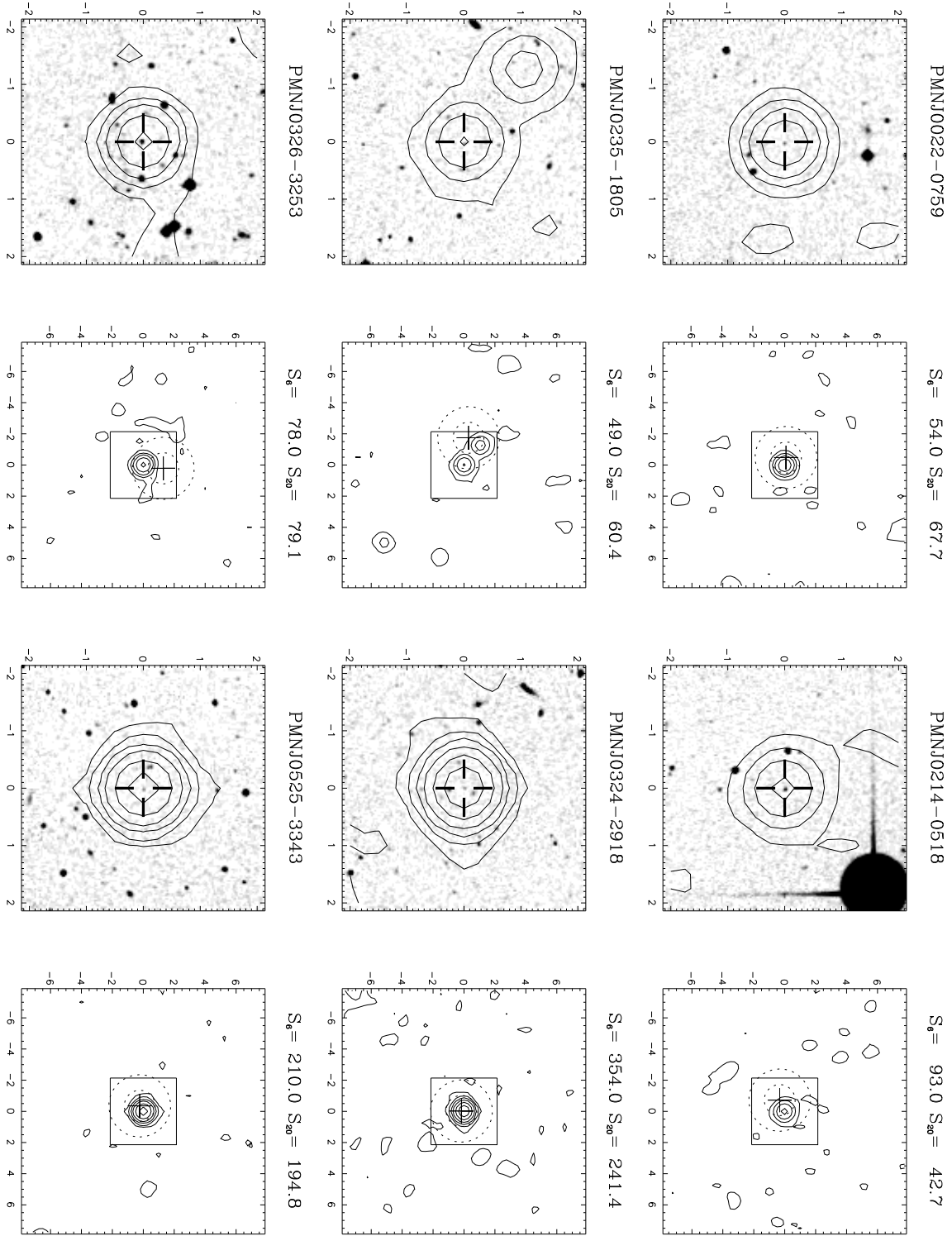,height=8.0in,bbllx=61pt,bblly=50pt,bburx=572pt,bbury=717pt}}
\caption{Left plot of each pair: NVSS Radio contours superimposed on the DSS
(blue) images. The cross marks the position of the APM identification
(often not visible in B). Right panel of each pair: Larger scale map,
16 arcmin across, showing the NVSS contours. The square shows the
field of the left hand panel (4 arcmin across). The cross and
concentric dotted circles show the PMN position and approximate beam
size. The outer circle corresponds to a 2 arcminute beam radius.
On both plots the contours are from the 1.4GHz NVSS maps and the
contour levels are $0.004*(2.0)^i-0.003$Jy, where i ranges from 0 to
11.}
\label{nvss}
\end{figure*}

\begin{figure*} 
\centerline{\psfig{figure=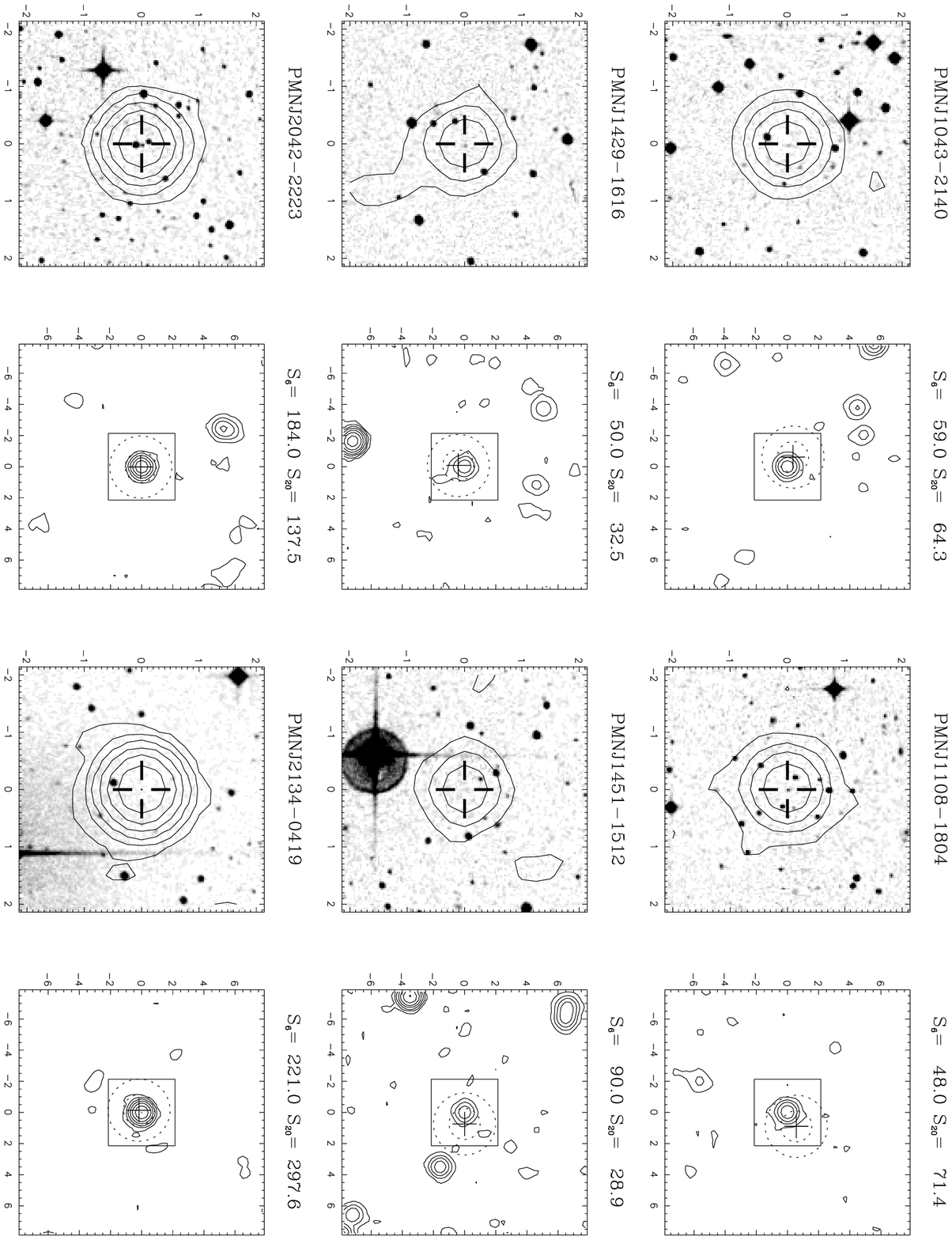,height=8.0in,bbllx=61pt,bblly=50pt,bburx=572pt,bbury=717pt}}
\end{figure*}

\begin{figure*} 
\centerline{\psfig{figure=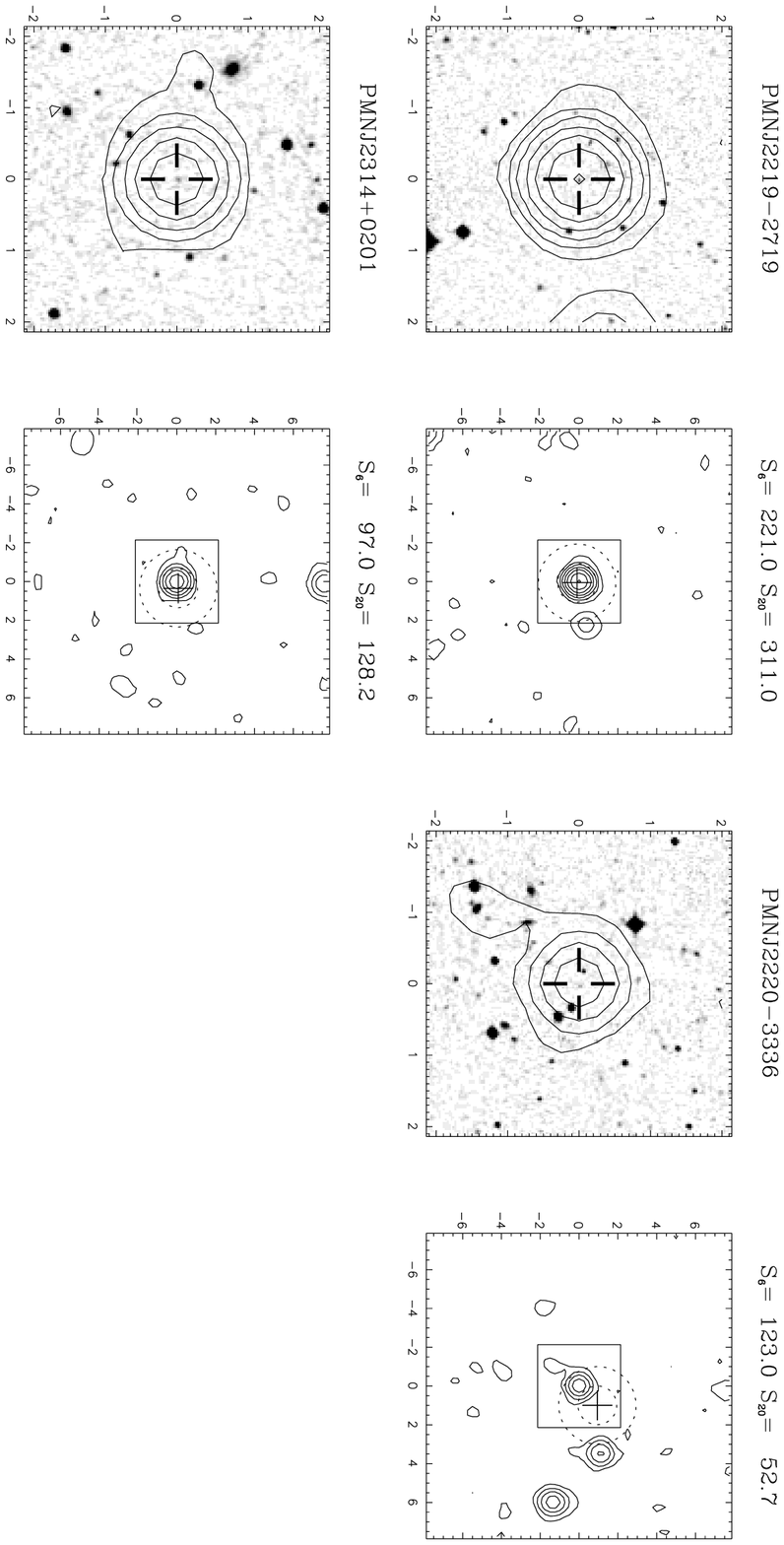,height=8.0in,bbllx=61pt,bblly=50pt,bburx=572pt,bbury=717pt}}
\end{figure*}

\section{Discussion}

By using radio and optical multicolour data covering a significant
fraction of the sky, we have produced a sample of high-redshift,
optically bright radio-loud quasars.

Since this new quasar sample is well-defined, we can use it to
estimate the surface density of $z>4$ quasars. To do this we consider
objects with radio flux densities above the brightest limit of the PMN
surveys, 72mJy. The SGC region of our survey has a high completeness
of spectroscopic follow up (only two objects, or 1\%, were not
observed in this region) and also had a high completeness of the NVSS
survey (0.955). There are four $z>4$ QSOs in the SGC region of the
complete sample, which implies a surface density of $\rm 0.92\pm
0.5\times 10^{-3} sq\ deg^{-1}$ for $z>4$ QSOs with $\rm S\ge 72mJy$.

If the whole survey is considered, and the completeness for the SGC
and NGC are taken as 94\% and 61\% respectively (as implied by the
number of spectroscopically observed candidates in each region
combined with the NVSS completeness factors derived in section 2) then
the derived surface density is $\rm 1.0\pm 0.4\times 10^{-3} sq\
deg^{-1}$. This is fully consistent with the value determined from the
SGC alone.

This value is similar to that of 1 per 1600sq deg ($6.3\times
10^{-4}$) found by Snellen et al (2001) using a similar technique
although with a slightly different radio and optical flux density
limits ($\rm R= 20$ compared to R=21 for the current survey, $\rm S\ge
30mJy$ compared to $\rm S>72mJy$, an upper redshift limit defined by
the red optical filter of $z\sim4.5$ rather than $\sim4.7$, and a
radio spectral index cut at $-0.35$ rather than $-0.5$).

There were no quasars found in our survey with $z>4.76$ despite the
fact that bright quasars with 4.9\lapprox {\it z}\lapprox 6.3 should
have been detectable on the I-plates. Allowing for the 86\%
completeness of spectroscopic follow up of the I-band sample, the
effective area covered was 2930 sq deg. Therefore we derive an upper
limit of $\rm 3.4\times 10^{-4} sq\ deg^{-1}$ for the surface density of
flat spectrum quasars with $\rm I<19.5$, $\rm S_{5GHz}\ge 25mJy$ and
4.9\lapprox {\it z}\lapprox 6.3.

Finally, the new sample of quasars presented in this paper represent
some of the most luminous objects in the Universe and may also
represent extreme peaks in the matter density distribution at
high-redshift.  They are therefore ideal targets for various follow-up
programs such as high-redshift absorption line studies and 
searches for associated high-redshift clusters.

\begin{acknowledgements}

RGM thanks the Royal Society for support.  We thank Jason Spyromilio
for providing the IR spectrum of PMN J1451-1512.

We also thank Mike Irwin and the staff at the APM facility in
Cambridge for producing the scans of UKST plates used in this survey.

This publication makes use of data products from the Two Micron All
Sky Survey, which is a joint project of the University of
Massachusetts and the Infrared Processing and Analysis Center, funded
by the National Aeronautics and Space Administration and the National
Science Foundation.

\end{acknowledgements}

\end{document}